\documentclass[12pt]{article}
\include{epsf}
\include{psfig}
\include{epsfig}
\usepackage{epsfig}
\makeatletter

\def\compoundrel#1\over#2{\mathpalette\compoundreL{{#1}\over{#2}}}
\def\compoundreL#1#2{\compoundREL#1#2}
\def\compoundREL#1#2\over#3{\mathrel
         {\vcenter{\hbox{$\m@th\buildrel{#1#2}\over{#1#3}$}}}}
\makeatother

\makeatletter

\begin{document}
\setcounter{page}{0}
\thispagestyle{empty}
\setlength{\parindent}{1.0em}
\begin{flushright}
GUTPA/03/12/01
\end{flushright}
\renewcommand{\thefootnote}{\fnsymbol{footnote}}
\begin{center}{\LARGE{{\bf Hierarchy Problem and a new Bound 
State\footnote{\it To be published in the Proceedings of the Euresco 
Conference on What comes beyond the Standard Model? Symmetries beyond 
the Standard Model, Portoroz, Slovenia, 12 - 17 July 2003.} }}}
\end{center}
\begin{center}{\large{C. D. Froggatt}
\\}
\end{center}
\renewcommand{\thefootnote}{\arabic{footnote}}
\begin{center}{{\it Department of Physics and Astronomy}\\{\it
University of Glasgow, Glasgow G12 8QQ, Scotland}}\end{center}
\begin{center}{\large{H. B.
Nielsen}
\\}
\end{center}
\renewcommand{\thefootnote}{\arabic{footnote}}
\begin{center}{{\it  Niels Bohr Institute, }\\{\it
Blegdamsvej 17-21, DK 2100 Copenhagen, Denmark}}\end{center}
\setcounter{footnote}{0}

\begin{abstract}
Instead of solving fine-tuning problems by some automatic method
or by cancelling the quadratic divergencies in the hierarchy
problem by a symmetry (such as SUSY), we rather propose to look
for a unification of the different fine-tuning problems. Our
unified fine-tuning postulate is the so-called Multiple Point
Principle, according to which there exist many vacuum states with
approximately the same energy density (i.e.~zero cosmological
constant). Our main point here is to suggest a scenario, using
only the pure Standard Model, in which an exponentially large
ratio of the electroweak scale to the Planck scale results. This
huge scale ratio occurs due to the required degeneracy of three
suggested vacuum states. The scenario is built on the hypothesis
that a bound state formed from 6 top quarks and 6 anti-top quarks,
held together mainly by Higgs particle exchange, is so strongly
bound that it can become tachyonic and condense in one of the
three suggested vacua. If we live in this vacuum, the new bound
state would be seen via its mixing with the Higgs particle. It
would have essentially the same decay branching ratios as a Higgs
particle of the same mass, but the total lifetime and production
rate would deviate from those of a genuine Higgs particle.
Possible effects on the $\rho$ parameter are discussed.
\end{abstract}
\thispagestyle{empty}

\newpage

\section{Introduction}
\label{introduction}

There are several problems in high energy physics and cosmology of
a fine-tuning nature, such as the cosmological constant problem or
the problem of why the electroweak scale is lower than the Planck
energy scale by a huge factor of the order of $10^{17}$. In
renormalisable theories, such fine-tuning problems reappear order
by order in perturbation theory. Divergent, or rather cut-off
dependent, contributions (diagrams) have to be compensated by
wildly different bare parameters order by order. The most well-known
example is the hierarchy problem in a non-supersymmetric theory
like the pure Standard Model, which is the model we consider in this
article. New quadratic divergencies occur order by order in the square
of the Standard Model Higgs mass, requiring the bare Higgs mass squared
to be fine-tuned again and again as the calculation proceeds order by
order. If, as we shall assume, the cut-off reflects
new physics entering near the Planck scale $\Lambda_{Planck}$,
these quadratic divergencies
become about $10^{34}$ times bigger than the final mass squared of
the Higgs particle. Clearly an explanation for such a fine-tuning
by ~34 digits is needed. Supersymmetry can tame these divergencies
by having a cancellation between fermion and boson contributions,
thereby solving the technical hierarchy problem. However the problem
of the origin of the huge scale ratio still remains, in the form of
understanding why the $\mu$-term and the soft supersymmetry breaking
terms are so small compared to the fundamental mass scale
$\Lambda_{Planck}$.

In addition to the fine-tuning problems, there are circa 20
parameters in the Standard Model characterising the couplings and
masses of the fundamental particles, whose values can only be
understood in speculative models extending the Standard Model. On
the other hand, the only direct evidence for physics beyond the
Standard Model comes from neutrino oscillations and various
cosmological and astrophysical phenomena. The latter allude to
dark matter, the baryon number asymmetry  and the need for an
inflaton field or some other physics to generate inflation. In
first approximation one might ignore such indications of new
physics and consider the possibility that the Standard Model
represents physics well, order of magnitudewise, up to the Planck
scale.

In the short term, rather than a new extended model with new
fields, we have the need for an extra principle that can specify
the values of the fine-tuned parameters and give predictions for
theoretically unknown parameters. Of course we do need new fields
or particles, such as dark matter, heavy see-saw neutrinos and the
inflaton, but at present they constitute a rather weak source of
inspiration for constructing the model beyond the Standard Model.
On the other hand there is a strong call for an understanding of
the parameters, e.g.~the cosmological constant or the Higgs
particle mass, in the already well working Standard Model.

Since these problems are {\em only} fine-tuning problems, it would
{\em a priori} seem that we should look for some fine-tuning principle.
In a renormalisable theory, a fine-tuning requirement should concern
renormalised parameters rather than bare ones. This is well
illustrated by the cosmological constant problem, where simply
requiring a small value for the bare cosmological constant will not
solve the phenomenological problem. There are several contributions
(e.g.~from electroweak symmetry breaking)
to the observed value, renormalising it so to speak, which are huge
compared to the phenomenological value.


\section{A Fine-tuning Principle}
\label{finetuning}

In the spirit of renormalisable theories, it is natural to
formulate a fine-tuning postulate in terms of quantities that are
at least in principle experimentally accessible. So one is led to
consider n-point functions or scattering amplitudes, which are
functions of the 4-momenta of the external particles. However, in
order to specify the value of such a quantity for a given
configuration of particles, it would be necessary to specify all
the external momenta of the proposed configuration. One might
consider taking some integral or some average over all the
external momenta in a clever way. However, for several external
particles, it really looks rather hard to invent a fine-tuning
postulate that is simple enough to serve as a fundamental
principle to be fulfilled by Nature in choosing the coupling
constants and masses. But the situation becomes much simpler if we
think of formulating a fine-tuning principle for a zero-point
function! The zero-point function is really just the vacuum energy
density or the value of the dressed cosmological constant
$\Lambda_{cosmo}$. The cosmological constant is of course a good
idea for our purpose, in as far as the cosmological constant
problem itself would come into the fine-tuning scheme immediately
if we make the postulate that the zero-point function should
vanish. Nowadays its fitted value is not precisely zero, in as far
as about $73\%$ of the energy density in the Universe is in the
form of dark energy or a cosmological constant. However this value
of the cosmological constant is anyway very tiny compared to the
{\em a priori} expected Planck scale or even compared to the
electroweak or QCD scales.

Now an interesting question arises concerning the detailed form of
this zero cosmological constant postulate, in the case when there
are several candidate vacuum states. One would then namely ask:
should the zero cosmological constant postulate apply just to one
possible vacuum state or should we postulate that all the
candidate vacua should have their {\em a priori} different
cosmological constants set equal to zero (approximately)? It is
our main point here to answer this question by extending the zero
cosmological constant postulate to all the candidate vacua! In
fact this form of the zero cosmological constant postulate
unifies\footnote{We thank L.~Susskind for pointing out to us that
the cosmological constant being zero can be naturally incorporated
into the Multiple Point Principle} the cosmological constant
problem with our so-called Multiple Point Principle
\cite{glasgowbrioni}, which states that there exist several vacua
having approximately the same energy density.

In principle, for each proposed method for explaining why the
cosmological constant is approximately zero, we can ask whether it
works for just the vacuum that is truly realised or whether it
will make several vacuum candidates zero by the same mechanism.
For example, one would expect that the proposal of Guendelman
\cite{guendelmann}, of using an unusual measure on space-time, would
indeed easily give several vacua with zero energy density rather
than only one. However for a method like that of Tsamis and
Woodard  \cite{woodard}, in which it is the actual time
development of the Universe that brings about the effectively zero
cosmological constant, one would only expect it to work in the
actual vacuum. Similarly if one uses the anthropic principle
\cite{weinberg}, one would only expect to get zero cosmological
constant for that vacuum in which we, the human beings, live.

The main point of the present talk is to emphasize that the
Multiple Point Principle, which can be considered as a consequence
of solving the cosmological constant problem in many vacua, can be
helpful in solving other fine-tuning problems; in particular the
problem of the electroweak scale being so tiny compared to the
Planck scale.

\section{Approaching the Large Scale Ratio Problem}
\label{scaleratio}

We consider here the problem of the hierarchy between the Planck
scale and the electroweak scale. This scale ratio is so huge that
it is natural to express it as the exponential of a large number.
In fact we might look for inspiration at another scale ratio
problem for which we already have a good explanation: the ratio of
the fundamental (Planck) scale to the QCD scale. The QCD scale
$\Lambda_{QCD}$ is the energy scale at which the QCD fine
structure constant formally diverges. It is believed that the
scale ratio $\Lambda_{Planck}/\Lambda_{QCD}$ is determined by the
renormalisation group running of the QCD fine structure constant
$\alpha_s(\mu)$, with the scale ratio being essentially equal to
the exponent of the inverse of the value of the fine structure
constant $1/\alpha_s(\Lambda_{Planck})$ at the Planck scale. So we
might anticipate explaining the Planck to electroweak scale ratio
in terms of the renormalisation group and the Standard Model
running coupling constants at the fundamental scale and the
electroweak scale.

At first sight, it looks difficult to get such an explanation by
fine-tuning a running coupling -- e.g.~the top quark Yukawa
coupling -- at the electroweak scale, using our requirement of
having vacua with degenerate energy densities. The difficulty is
that, from simple dimensional arguments, the energy density or
cosmological constant tends to become dominated by the very
highest frequencies and wave numbers relevant in the quantum field
theory under consideration -- the Planck scale in our case. In
fact the energy density has the dimension of energy to the fourth
power, so that modes with Planck scale frequencies contribute
typically $(10^{17})^4$ times more than those at the electroweak
scale.  The only hope of having any sensitivity to electroweak
scale physics would, therefore, seem to be the existence of two
degenerate phases, which are identical with respect to the state
of all the modes corresponding to higher than electroweak scale
frequencies. They should, so to speak, \underline{only} deviate by
their physics at the electroweak scale and perhaps at lower scales
in energy. In such a case it could be that the energy density
difference between the two phases would only depend on the
electroweak scale physics and, thus, could more easily depend on
the running couplings taken at the electroweak scale. It is,
namely, only for the modes of this electroweak scale that the
running couplings at this scale are relevant.

So, in order to ``solve'' the large scale ratio problem using our
Multiple Point Principle, we need to have a model with two
different phases that only deviate by the physics at the
electroweak scale. So what could that now be? Different phases are
most easily obtained by having different expectation values of
some scalar field, which really means different amounts of some
Bose-Einstein condensate. A nice way to have such a condensate only 
involve physics at a certain low energy scale, the electroweak scale 
say, consists in having a condensate of bound states made out of some
Standard Model particles -- we shall actually propose top quarks
and anti-top quarks. Such bound states could now naturally have
sizes of the order of the electroweak length scale. Such a picture
would really only make intuitive sense, when the density is not
large compared to the scale given by the size of the bound states;
otherwise they would lie on top of each other and completely
disturb the binding. One might say that the physical situation for
the binding would become drastically changed, when the density in
the bound state condensate gets so high as to have huge multiple
overlap. Presumably one could naturally get a condensate with a
density which is not so far from the scale given by the size and,
thus, the electroweak scale.  In the next section we shall spell
out this idea of making a bound state condensate in more detail.
We shall then return to the large scale ratio problem in section
\ref{return} and explain how the Multiple Point Principle is used
to determine the top quark Yukawa coupling constant at the Planck
scale, in terms of the electroweak gauge coupling constants, by
postulating the existence of a third degenerate vacuum.

\section{The Bound State}
\label{boundstate}

\subsection{The Idea of a Bound State Condensate}
\label{condensate}

So we are led to consider some strongly bound states, made out of
e.g. top-quarks and using Higgs fields or other particles to bind
them, such that the energy scale of a condensate formed from them
is - by dimensional arguments - connected to the scale of the
Standard Model Higgs field vacuum expectation value or VEV (which
is of course what one usually calls the electroweak scale). For
dimensional reasons this condensate has now a density of an order
of magnitude given by this electroweak scale. Then the frequencies
or energies of the involved modes of vibration are also of this
order, in the sense that it is the modes with energies of this
order that make the difference (between two phases say).  It is
therefore also the running couplings at this scale that are the
directly relevant parameters! If we now impose some condition,
like the degeneracy of two phases resulting from this bound state
condensation dynamics, it should result in some relation or
requirement concerning the {\em running couplings at the electroweak
scale}.

Instead of simply a bound state condensing, one could {\em a priori}
also hope for some other nonlinear effect taking place in a way
involving essentially only the modes/physics at the electroweak
scale. The crux of the matter is that, at short distances compared
to the electroweak length scale, the non-perturbative effect in
question would hardly be felt. Consequently, the huge contributions
to the energy density from the short distance modes can be cancelled
out between the two phases, in imposing our Multiple Point Principle
(MPP). But the bound state idea is in a way the most natural and
simple, since bound states are already well-known to occur in many 
places in quantum physics.

\subsection{A Bound State of 6 top and 6 anti-top quarks?}
\label{12tops}

Of course, when we look for bound states in the Standard Model, we
know immediately that there is a huge number of hadronic bound
states consisting of mesons, baryons and glueballs, i.e. from QCD.
These bound states typically have the size given by the QCD
scale parameter $\Lambda_{QCD}$, which means lengths of the order 
of an inverse GeV. That is to say the strong scale rather than the
electroweak one. Nevertheless you could {\em a priori} hope that
some phase transition, involving quarks and caused by QCD, could
determine a certain quark mass by the MPP requirement of being at
the border between two phases of the vacuum. That could then in
turn lead to a fixing of the Standard Model Higgs VEV, which is
known to be responsible for all the quark masses. But, from
dimensional considerations, you would expect to get that the quark
mass needed by such an MPP mechanism would be of the order of the
strong scale. So it may work this way if the strange quark were
the one to be used, since it namely has a mass of the order of the
strong scale. But these speculated QCD-caused phase transitions
are not quite what we require. We rather seek a condensation
getting its scale from the Standard Model Higgs VEV and want to
avoid making severe use of QCD. However, at first sight, the other
gauge couplings and even the top quark Yukawa coupling (let alone
the other smaller Yukawa couplings) seem rather small for making
strongly bound states; with a binding so strong, in fact, as to
make the bound state tachyonic and to condense in the vacuum.
Indeed, if you think of bound states consisting of a couple of
particles, it is really pretty hopeless to find any case of such a
strong binding except in QCD. But now scalar particle exchange has
an important special feature. Unlike the exchange of gauge
particles, which lead to alternating signs of the interaction when
many constituents are put together, scalar particle exchange leads
to attraction in all cases: particles attract both other particles
and antiparticles and the attraction of quarks, say by Higgs
exchange, is independent of colour.

The only hope of getting very strong binding without using QCD, so
as to obtain tachyonic bound states in the Standard Model, is to
have many particles bound together and acting cooperatively -- and
then practically the use of a scalar exchange is unavoidable. So we
are driven towards looking for bound states caused dominantly by the
exchange of Higgs particles, since the Higgs particle is the only
scalar in the Standard Model and we take the attitude of
minimising the amount of new physics. Since the Yukawa couplings
of the other quarks are so small, our suggestion is to imagine
some top quarks and/or anti-top quarks binding together into an
exotic meson. It better be a boson and thus a ``meson'', since we
want it to condense.

There are, of course, bound states of say a top quark and an
anti-top quark which are mainly bound by gluon exchange, although
comparably by Higgs exchange. However these are rather loosely
bound resonances compared to the top quark mass. But, if we now add more
top or anti-top quarks to such a state, the Higgs exchange
continues to attract while the gluon exchange saturates and
gets less significant. This means that the Higgs exchange binding
potential for the whole system gets proportional to the number of
pairs of constituents, rather than to the number of constituents
itself. So, at least {\em a priori} by having sufficiently many
constituents, one might foresee the binding energy exceeding the
constituent mass of the system.

In order to get the maximal binding, one needs to put each of the
added quarks or anti-quarks into an S-wave state. Basically we can
use the same technique as in the calculation of the binding energy
of the electron to the atomic nucleus in the hydrogen atom. Once
the S-wave states are filled, we must go over to the P-wave and so
on. But the P-wave binding in a Coulomb shaped potential only
provides one quarter of the binding energy of the S-wave. In the
case of a scalar exchange, an added particle gets a binding energy
to each of the other particles already there. However, once the
S-wave states are filled, these binding energies go down by a
factor of about four in strength and it becomes less profitable
energetically to add another particle. Depending on the strength
of the coupling, it can therefore very easily turn out to be most
profitable to fill the S-wave states and then stop.

Now, when we use top quarks and anti-top quarks, one can easily
count the number of constituents, by thinking of the S-wave as
meaning that essentially all the particles are in the same state
in geometrical space. Then there are 12 different ``internal''
states into which these S-wave quarks/antiquarks can go: each
quark or anti-quark can be in two spin states and three colour
states, making up all together $2 \times 2 \times 3 = 12$ states.
Thus we can have 6 top quarks and 6 anti-top quarks in the bound
state, before it gets necessary to use the P-wave. We shall here
make the hypothesis, which to some extent we check below, that
indeed the strongly bound state which we seek is precisely this
one consisting of just 12 particles.

So we now turn to the question of whether or not this exotic 6 top
quark and 6 anti-top quark state is bound sufficiently strongly to
become tachyonic, i.e.~to get a negative mass squared. Actually,
in order to confirm our proposed MPP fine-tuning mechanism, we
need that the experimentally measured top quark Yukawa coupling
should coincide with the borderline value between a condensate of
this almost tachyonic exotic meson being formed or not being
formed. On the basis of the following crude estimate, we want to
claim that such a coincidence is indeed not excluded.

\subsection{The Binding Energy Estimate}
\label{binding_estimate}

We now make a crude estimate of the binding energy of the proposed
12 quark/anti-quark bound state. As a first step we consider the
binding energy $E_1$ of one of them to the remaining 11
constituents treated as just one collective particle, analogous to
the nucleus in the hydrogen atom. Provided that the radius of the
system turns out to be sufficiently small compared to the Compton
wavelength of the Standard Model Higgs particle, we can take this
to be given by the well-known Bohr formula for the ground state
binding energy of a one electron atom. It is simply necessary to
replace the electric charge of the electron $e$ by the top quark
Yukawa coupling $g_t/\sqrt{2}$, in the normalisation where the
running mass of the top quark is given by the formula $m_t = g_t
\, 174$ GeV, and to take the atomic number to be $Z=11$:
\begin{equation}
E_1 = -\left(\frac{11g_t^2/2}{4\pi}\right)^2 \frac{11m_t}{24}
\label{binding}
\end{equation}
Here we have used $m_t^{reduced}= 11m_t/12$ as the reduced mass of
the top quark.

In order to obtain the full binding energy for the 12 particle
system, we should multiply the above expression by 12 and divide
by 2 to avoid double-counting the pairwise binding contributions.
However this analogy with the atomic system only takes into
account the $t$-channel exchange of a Higgs particle between the
constituents. A simple estimate of the $u$-channel Higgs exchange
contribution \cite{itep} increases the binding energy by a further
factor of $(16/11)^2$. So the expression for the total non-relativistic
binding energy due to Higgs particle exchange interactions
becomes:
\begin{equation}
 E_{binding} = \left(\frac{11g_t^4}{\pi^2}\right)m_t
\label{binding2}
\end{equation}

We have here neglected the attraction due to gluon exchange and
the even smaller electroweak gauge field forces. However the gluon
attraction is rather a small effect compared to the Higgs particle
exchange, in spite of the fact that the QCD coupling
$\alpha_s(M_Z) = g_s^2(M_Z)/4\pi = 0.118$. This value of the QCD
fine structure constant corresponds to an effective gluon top
anti-top coupling constant squared of:
\begin{equation}
e_{tt}^2 = \frac{4}{3}g_s^2 \simeq \frac{4}{3} 1.5 \simeq 2.0
\end{equation}
We have to compare this gluon coupling strength $e_{tt}^2 \simeq
2$ with $Zg_t^2/2 \simeq 11/2 \times 1.0$ from the Higgs particle.
This leads to an increase of the binding energy by a factor of
$(15/11)^2$ due to gluon exchange, giving our final result for the
non-relativistic binding energy:
\begin{equation}
 E_{binding} = \left(\frac{225g_t^4}{11\pi^2}\right)m_t
\label{binding3}
\end{equation}

The correction from W-exchange will be smaller than that from
gluon exchange by a multiplicative factor of about
$\left(\frac{\alpha_2(M_Z)}{\alpha_s(M_Z)} \frac{3}{4}\right)^2 \simeq
\frac{1}{25}$, and the weak hypercharge exchange is further
reduced by a factor of $\sin^4\theta_W$. Also the s-channel Higgs
exchange diagrams will give a contribution in the same direction.
There are however several effects going in the opposite direction,
such as the Higgs particle not being truly massless and that we
have over-estimated the concentration of the 11 constituents
forming the ``nucleus''. Furthermore we should consider
relativistic corrections, but we postpone a discussion of their
effects to ref.~\cite{fln}.

\subsection{Estimation of Phase Transition Coupling}
\label{coupling_estimate}

From consideration of a series of Feynman diagrams or the
Bethe-Salpeter equation for the 12 particle bound state, we would
expect that the mass squared of the bound state, $m_{bound}^2$,
should be a more analytic function of $g_t^2$ than $m_{bound}$
itself. So we now write a Taylor expansion in
$g_t^2$ for the mass {\em squared} of the bound state, crudely
estimated from our non-relativistic binding energy formula:
\begin{eqnarray}
m_{bound}^2 & = & \left(12m_t\right)^2 - 2\left(12
m_t\right)\times
E_{binding} + ...\\
& = & \left(12m_t\right)^2\left(1 -\frac{225}{66\pi^2}g_t^4 +
...\right)
\label{expansion}
\end{eqnarray}

We now assume that, to first approximation, the above formal
Taylor expansion (\ref{expansion}) can be trusted even for large
$g_t$ and with the neglect of higher order terms in the {\em mass
squared} of the bound state. Then the condition that the bound
state should become tachyonic, $m_{bound}^2 < 0$, is that the top
quark Yukawa coupling should be greater than the value given by
the vanishing of equation (\ref{expansion}):
\begin{equation}
0 =  1-\frac{225}{66\pi^2}g_t^4 + ...
\label{br0}
\end{equation}
We expect that once the bound state becomes a tachyon, we should
be in a vacuum state in which the effective field, $\phi_{bound}$,
describing the bound state has a non-zero expectation value. Thus
we expect a phase transition just when the bound state mass
squared passes zero\footnote{In fact the phase transition
(degenerate vacuum condition) could easily occur for a small
positive value of $m_{bound}$ and hence a somewhat smaller value
of $g_t^2$.}, which roughly occurs when the running top quark
Yukawa coupling at the electroweak scale, $g_t(\mu_{weak})$,
satisfies the condition (\ref{br0}) or:
\begin{equation}
g_t|_{phase \ transition} =
\left(\frac{66\pi^2}{225}\right)^{1/4} \simeq 1.3
\end{equation}

We can make an estimate of one source of uncertainty, by
considering the effect of using a leading order Taylor expansion
in $g_t^2$ for $m_{bound}$ instead of for $m_{bound}^2$. This
would have led to difference of a factor of 2 in the binding
strength and hence a correction by a factor of the fourth root of
2 in the top quark Yukawa coupling at the phase boundary; this
means a $20\%$ uncertainty in $g_t|_{phase \ transition}$. Within
an uncertainty of this order of $20\%$, we have a 1.5 standard
deviation difference between the phase transition (and thus the
MPP predicted) coupling, $g_t \simeq 1.3$, and the measured one,
$g_t \simeq 1.0$, corresponding to a physical top quark mass of
about 173 GeV. We thus see that it is quite conceivable within our
very crude calculations that, with the experimental value of the
top quark Yukawa coupling constant, the pure Standard Model could
lie on the boundary to a new phase; this phase is characterised by
a Bose-Einstein condensate of bound states of the described type,
consisting of 6 top quarks and 6 anti-top quarks!

\subsection{Mixing between the Bound State and the Higgs Particle}

Strictly speaking, if the above scenario is correct, it is not at
all obvious in which of the two vacua we live. If we live in the
phase in which the bound state condensate is present, the
interaction of the bound state particle with the Standard Model
Higgs particle can cause a bound state particle to be pulled out
of the vacuum condensate and then to function as a normal
particle. This effect will mean that the normal Higgs particle
will {\em mix} with the bound state, in a similar way as one has
mixing between the photon and the $Z^0$ gauge boson, or between
$\eta$ and $\eta'$. This means that the two observed particles
would actually be superpositions, each with some amplitude for
being the bound state and with some amplitude for being the
original Higgs particle. Both can have expectation values, or
rather the expectation value is described by some abstract vector
denoting the two different components. Also both superpositions
would be exchanged and contribute to the binding of the bound
state. Taking this two component nature of the effective Higgs
particles into account makes the discussion more complicated than
with a single Higgs particle.

Really there are three types of experimentally accessible parameters
for which we at first want to predict a relation from our bound
state model:
\begin{enumerate}
\item
The top quark mass is given in the simplest case by the top quark
Yukawa coupling times the Higgs VEV. However, in the two effective
Higgs picture (one of them being the bound state essentially just
mixed somewhat with the Higgs particle), the top quark mass
becomes of the form $h_1 v_1 + h_2 v_2$. Here $h_1$ and $h_2$ are
the Yukawa couplings of the top quark for the two superpositions,
whose VEVs are denoted by $v_1$ and $v_2$.
\item
The gauge boson masses: for example the $W$ boson mass in the
two effective Higgs picture becomes $M_{W}^2= g_2^2 (v_1^2 + v_2^2)$.
Here, for simplicity, we have taken both the fields to be doublets
with weak hypercharge $y/2= -1/2$ like the original Higgs field.
We reconsider the irreducible representation content of the bound
state field in section \ref{rhoparameter}, where we discuss the
$\rho$ parameter problem.
\item
The binding strength parameter for the bound state which
determines the vacuum phase in which the energy density is the
lowest. Even if this parameter is hard to determine
experimentally, we may at least relate it to our MPP from which it
can essentially be predicted. In the simplest case with a single
Higgs particle, the binding strength parameter is just the top
quark Yukawa coupling squared $g_t^2$ as discussed in section
\ref{coupling_estimate}. However with two effective Higgs
particles, the parameter $g_t^2$ would to first approximation be
replaced by $ h_1^2 + h_2^2$.
\end{enumerate}

We now remark that the three quantities listed above
are related by a Schwarz inequality, namely:
\begin{equation}
|h_1v_1 + h_2v_2|^2 \le (v_1^2 + v_2^2)(h_1^2 + h_2^2)
\end{equation}
(written as if we had only real numbers, but we could use complex
ones also). With this correction due to the mixing, we lose our
strict prediction of the top quark mass corresponding to two
degenerate phases, with and without a bound state condensate
respectively, unless we can estimate the mixing. In fact such an
estimate is not entirely out of question, because we know the
coupling of the Higgs to the bound state and can potentially also
estimate the density of the condensate. Qualitatively we just
predict that the resulting top quark mass will be somewhat {\em
smaller} than the estimate made in section
\ref{coupling_estimate}, but we expect it to remain of a similar
order of magnitude.

Such a mixing correction would seem to be welcome, in order to improve
the agreement of the experimental top quark Yukawa coupling with
the estimated phase transition value. We namely tend to predict
a top quark mass, which is too large by a factor of about 1.3, without
including any mixing correction. However this disagreement should not be
taken too seriously, as it is within the accuracy of our calculation.
Nonetheless, if we do live in the phase containing the bound state
condensate, the mixing correction would be good for repairing this
weak disagreement with experiment.

\section{Return to the Large Scale Ratio Problem}
\label{return}

\subsection{Three degenerate vacua in the pure Standard Model}

As discussed at Portoroz by Colin Froggatt \cite{colin}, it is
possible to determine the value of the top quark running Yukawa 
coupling $g_t(\mu)$ at the fundamental scale $\mu_{fundamental} =
\Lambda_{Planck}$ by using the Multiple Point Principle to
postulate the existence of a third degenerate vacuum, in which the
Standard Model Higgs field has a VEV of order the Planck scale
\cite{fn2}. This requires that the renormalisation group improved
effective potential for the Standard Model Higgs field should have
a second minimum near the Planck scale, where the potential should
essentially vanish. This in turn means that the Higgs
self-coupling constant $\lambda(\mu)$ and its beta function
$\beta_{\lambda}(\mu)$ should both vanish near the fundamental
scale, giving the following relationship between the top quark
Yukawa coupling $g_t(\mu_{fundamental})$ and the electroweak
$SU(2) \times U(1)$ gauge coupling constants
$g_2(\mu_{fundamental})$ and $g_1(\mu_{fundamental})$:
\begin{equation}
\label{gt4} g_t^4 = \frac{1}{48} \left(9g_2^4 + 6g_2^2g_1^2
+3g_1^4 \right)
\end{equation}
If we now input the experimental values of the gauge coupling
constants, extrapolated to the Planck scale using the Standard
Model renormalisation group equations, we obtain
$g_t(\mu_{fundamental}) \simeq 0.4$. However we note that the
numerical value of $g_t(\mu)$, determined from the expression on
the right hand side of eq.~(\ref{gt4}), is rather insensitive to
the scale, varying by approximately $10\%$ between $\mu = 246$ GeV
and $\mu = 10^{19}$ GeV.

From our assumption of the existence of three degenerate vacua in
the Standard Model, our Multiple Point Principle has provided
predictions for the values of the top quark Yukawa coupling
constant at the electroweak scale, $g_t(\mu_{weak}) \simeq 1.3$,
and at the fundamental scale, $g_t(\mu_{fundamental}) \simeq 0.4$.
So we can now calculate a Multiple Point Principle prediction for
the ratio of these scales $\mu_{fundamental}/\mu_{weak}$, using
the Standard Model renormalisation group equations.

\subsection{Estimation of the logarithm of the scale ratio}

We now estimate the fundamental to electroweak scale ratio by
using the leading order beta function for the Standard Model top
Yukawa coupling constant $g_t(\mu)$:
\begin{equation}
 \beta_{g_t} = \frac{dg_t}{d\ln\mu} =
 \frac{g_t}{16\pi^2}\left(\frac{9}{2}g_t^2 - 8g_3^2 -
 \frac{9}{4}g_2^2 - \frac{17}{12}g_1^2\right)
 \label{betatop}
\end{equation}
where the $SU(3) \times SU(2) \times U(1)$ gauge coupling
constants are considered as given at the fundamental scale,
$\mu_{fundamental} = \Lambda_{Planck}$. It should be noticed that,
due to the relative smallness of the fine structure constants
$\alpha_i =g_i^2/4\pi$ and particularly of
$\alpha_3(\mu_{fundamental})$, the
beta function $\beta_{g_t}$ is numerically rather small at the
Planck scale. So the logarithm of the scale ratio
$\ln\mu_{fundamental}/\mu_{weak}$ needed to generate the required
amount of renormalisation group running, between the values
$g_t(\mu_{fundamental}) \simeq 0.4$ and $g_t(\mu_{weak}) \simeq
1.3$, must be a large number. Hence the scale ratio itself must be
huge and in this way we explain why the electroweak scale
$\mu_{weak}$ is so low compared to the fundamental scale
$\mu_{fundamental}$. In practice the Multiple Point Principle only
gives the order of magnitude of the logarithm of the scale ratio,
predicting $\mu_{fundamental}/\mu_{weak} \sim 10^{16} - 10^{20}$.

We note that as the strong scale is approached, $\mu \rightarrow
\Lambda_{QCD}$, $g_3(\mu)$ and the rate of logarithmic running of
$g_t(\mu)$ becomes large. So the strong scale $\Lambda_{QCD}$
provides an upper limit to the scale ratio predicted by the
Multiple Point Principle. Indeed the predicted ratio naturally
tends to give an electroweak scale within a few orders of
magnitude from the strong scale.

\section{How to see the bound state?}
\label{how-to-see}

\subsection{Mixing with the Higgs Particle}

Such a strongly bound state as we propose, consisting of 12
constituents, will practically act as a conserved type of
particle, because energy conservation forbids its destruction by
having a few of its constituents decay. The point is that the mass
of the remaining bound state or resonance, made up of the leftover
constituents, would be larger than that of the original strongly
bound state. Considering its interaction with the relatively light
particles of the Standard Model, the bound state would therefore
still be present after the interaction. This means that the most
important effective couplings, involving an effective scalar bound
state field and Standard Model fields, would have two or four
external bound state attachments. If we further restrict ourselves
to a renormalisable effective theory, we would be left with the
bound state scalar field only having interactions involving scalar
and gauge fields. An interaction between two fermions and two
scalar fields would already make up a dimension five operator,
which is non-renormalisable.

If we live in the phase without the condensate of new bound state
particles, these considerations imply that the bound state must be
long lived; it could only decay into a channel in which all 12
constituents disappeared together.  The production cross section
for such a particle would also be expected to be very low, if it
were just crudely related to the cross section for producing 6 top
quarks and 6 anti-top quarks. However, if we live in the phase
with the condensate, there exists the possibility that the bound
state particle could disappear into the condensate, which has of
course an uncertain number of bound state particles in it. Since,
as is readily seen, there is a significant coupling of the
Standard Model Higgs particle to two bound state attachments - a
three scalar coupling vertex - we can achieve such a disappearance
very easily by means of this vertex. This disappearance results in
the bound state obtaining an effective transition mass term into
the Standard Model Higgs particle. Such a transition means that
the Higgs particle and the new bound state will - provided we are
in the condensate phase - mix with each other! This has very
important consequences for the observability of the bound state.
We shall seemingly get two Higgs particles sharing the strength of
the fundamental Higgs particle, by each being a superposition of
the latter and of the bound state.

So, provided that we presently live in the phase with the bound
state condensate, we predict that at the LHC we shall apparently see
two Higgs particles! They will each behave just like the normal
Higgs particle, except that all of its couplings will be reduced
by a mixing angle factor, common of course for all the different
decay modes of the usual Standard Model Higgs particle, but
different for the two observed Higgs particles.

\subsection{The Rho Parameter; a Problem?}
\label{rhoparameter}

If we do not live in the phase with the condensate we can
naturally not expect to observe any effects of this condensate,
but if we live in the phase with the condensate then one might
look for the effects of this condensate. An effect that at first
seems to be there, and is perhaps likely to prevent the model from
being phenomenologically viable, is that the condensate of bound
states is not invariant under the $SU(2)\times U(1)$ gauge group
for the electroweak interactions. In fact this condensate will
{\em  a priori} begin to ``help'' the Standard Model Higgs field
giving masses to the $W$ and $Z^0$ particles. Now, however, since
we imagine the bound state to exist in the background of the usual
Higgs condensate and as only being bound due to the effects of
this surrounding medium, the bound state is strongly influenced by
the $SU(2)\times U(1)$ breaking effects of these surroundings.
Thus we cannot at first consider the bound state as belonging to
any definite irreducible representation of this electroweak group.
Rather we must either describe it by a series of fields belonging
to different irreducible representations of this group or simply
describe it by a single effective field that does not have any
definite electroweak quantum numbers. But this fact means that the
condensate of bound states has to be expressed by several such
fields having non-zero expectation values. These different fields
of different irreducible representations will not give the same
mass ratio for the $W$ and the $Z^0$ bosons. Thus, provided the
bound state condensate is of such an order of magnitude that its
effect on the gauge boson masses is not negligible, it will in 
general generate a $\rho$ parameter in disagreement with experiment.

So far our calculations have not supported the hope that, by some
mathematical accident, the $\rho$ parameter comes out to be
essentially equal to unity. Rather it seems that, in order for our
model to be consistent with the remarkably good agreement of the
Standard Model predictions with experiment, we require one of the
following situations to occur:

1) We do not live in the phase with the condensate but rather in
the one without the condensate.

2) The contribution of the condensate expectation value to the
gauge boson masses is simply very much smaller than that of the
genuine Standard Model Higgs field.

3) For irreducible representations other than the singlet and the
doublet with weak hypercharge $y/2 =1/2$, some self-interaction
or renormalisation group effect has made the irreducible
representation content of the bound state very small or vanishing.

As we shall see below, there is some weak evidence that our model
favours the idea that we actually live in the phase {\em with} the 
bound state condensate and even with an appreciable expectation value
compared to that of the genuine Higgs condensate. So it would seem
to fit our model best, if we could get the third of the above
possibilities to work and thereby avoid causing problems for the value of
the $\rho$ parameter.

\section{In which phase do \underline{we} live?}

In a model like ours, where there are many vacuum states, one must
identify which of those states is the vacuum around {\em us}. We
definitely live in a phase with a remarkably small Higgs field VEV
compared to the ``fundamental'' scale or natural unit for Higgs
field VEVs, which we take to be the Planck scale in our model.
Among the three Standard Model vacua discussed above, there are
thus two possibilities corresponding to the phases with the low
value of the Standard Model Higgs VEV. So what remains to be
decided is whether or not there is a condensate of the bound
states in the vacuum in which we live.

In section \ref{how-to-see} we discussed possible observational
effects related to the bound state, which could discriminate
between the two phases. Here we shall investigate which vacuum
phase is likely to emerge from the Big Bang and then assume that
it survives to the present epoch.

There is, however, no {\em a priori} reason to believe in the
absence of vacuum phase transitions since the first minutes after
the Big Bang. They might even have occurred in the era when stars
and galaxies were already present, but then one could imagine that
there should be astrophysical signatures revealing such transitions.
Indeed one might even wonder if the claims for a time variation of
the fine structure constant, indicated by some spectral investigations,
could be a consequence of such phase transitions. But it must be
admitted that the domain walls between phases would have such a
huge energy per unit surface area that they might be expected to
disturb all of cosmology as we understand it. So it seems likely that
there were no later phase transitions and that we do live in the
phase that emerged after the first minutes of the Big Bang.
If the other vacuum phases are to occur anywhere or
anytime at all, it must then be in the future.

We now turn to the question: what phase is likely to have come out
of the Big Bang? Of course the phase that emerges depends very
sensitively on the vacuum energy density. The higher energy
density vacua are expected to decay into the one with the lowest
energy density, provided though that sufficient thermal energy is
present to surmount any energy barriers between the vacua. Hence
the question of which vacuum emerges will be settled at the epoch
when the temperature is still just high enough that the phase
border can be passed, i.e.~when it is still possible to produce
the walls between the phases by thermal fluctuations.

At that epoch it is the Helmholz free energy density $f$ rather
than the true energy density $u$ that matters. The difference is
the term $-sT$, where $T$ is the temperature and $s$ is the
entropy density. Assuming the true energy density is exactly the
same in the two phases, the emergent phase should be the one
having the highest entropy density $s$ at the temperature in
question. That in turn should be the phase with highest number of
light species. Now, in the Standard Model, the known fermions and
gauge bosons, $W$ and $Z^0$, get their masses from the Higgs VEV
in the vacuum in question. So the emergent phase should 
be the one with the lowest Higgs VEV, when these 
particles have the smallest masses, giving in turn the larger
entropy, and then the lower free energy density. Now the presence
of the many bound states in the condensate tends to reduce the
Higgs field VEV. So it is indeed the phase {\em with} the
condensate, which is expected to come out from the early Universe.
Our tentative conclusion is thus that we should live in the phase
with the bound state condensate, provided of course that we are
correct in assuming that a new phase did not take over at a later
epoch.

Phenomenologically this phase with the bound state condensate present
today is the more interesting possibility, in as far as it leads to
the mixing of the Higgs particle and the bound state. It thereby
gives us the possibility of seeing this bound state much more easily,
namely as another ``Higgs'' particle. However, in this case we must
face up to the challenge of calculating the $\rho$ parameter.

\section{Conclusion}

In this talk, we have put forward a scenario for how the huge
hierarchy in energy scale comes about between a supposed
fundamental scale, taken as the Planck scale, and the electroweak
scale, meaning the scale of the $W$ and $Z^0$ particles and the
Higgs particle etc. This consists of introducing a fine-tuning
postulate -- the Multiple Point Principle -- according to which
there are many different vacua, in each of which the cosmological
constant or energy density is very small. In this way our main
fundamental assumption is that the cosmological constant problem
is solved, in some way or other, {\em several times}. The
remarkable result of the present article is that, as well as
fine-tuning the cosmological constants, this principle can lead to
a solution of a separate mystery, namely of why the electroweak
scale of energy is so low compared to the Planck scale. This
problem, which is separate from but closely related to the
technical hierarchy problem, gets solved in our scenario to the
degree that we even obtain a crude value for the logarithm of the
large scale ratio. We even get a suggestive explanation for why,
compared to its logarithmic distance from the Planck scale, the
electroweak scale is relatively close to the strong scale,
$\Lambda_{QCD}$. We, of course, have to input the large ratio 
of the Planck to QCD scales, in the form of the value of the 
QCD coupling constant at the fundamental scale.

In our scenario the pure Standard Model is assumed to be valid up
close to the Planck scale, apart from a possible minor
modification at the neutrino see-saw scale. We then postulate that
there are just three vacuum states all having, to first
approximation, zero energy density. In addition to specifying
information about the bare cosmological constant, this postulate
leads to two more restrictions between the parameters of the 
Standard Model. They are, in principle, complicated relations
between all the coupling constants and masses and it is
non-trivial to evaluate their consequences.  However we took the
values of the gauge coupling constants, which are anyway less
crucial, from experiment and these two relations then gave values
for the top quark Yukawa coupling constant at the
electroweak scale and the ``fundamental'' scale respectively:
$g_t(\mu_{weak}) \simeq 1.3$ and $g_t(\mu_{fundamental}) \simeq
0.4$.

The main point then is that we need an appreciable running of the
top quark Yukawa coupling, in order to make the two different
values compatible. That is to say we need a huge scale ratio,
since the running is rather slow due to the smallness of the
Standard Model coupling constants in general from the
renormalisation group point of view. This is our suggested
explanation for the mysterious huge hierarchy found empirically
between the Planck and electroweak scales. Indeed it even leads to
an approximately correct value for the logarithm of the huge scale
ratio!

It is crucial for our scenario that there should exist the
possibility of a phase with a certain bound state condensing in
the vacuum. The existence of such a bound state is {\em a priori}
a purely calculational problem, in which no fundamentally new
physics comes in. We suggest that this bound state should be
composed of 6 top quarks and 6 anti-top quarks held together by
Higgs exchange and, maybe to some extent, also by the exchange of
the bound state itself -- in a bootstrap-like way. If we live in
the vacuum without a bound state condensate, it would be difficult
to obtain direct experimental evidence for the bound state.
However if we live in the vacuum with a bound state condensate, 
which actually seems to be the
most likely situation in our scenario, it should be possible to
see the effects of this condensate. There should then be a
significant mixing between the bound state and the Higgs particle.
This implies the existence of two physical particles, sharing the
coupling strength and having the same decay branching ratios as
the conventional Standard Model Higgs particle. The resulting
effective 2 Higgs doublet model deviates from supersymmetry 
inspired models, by both ``Higgs"
particles having the same ratio of the couplings to the $-1/3$
charged quarks and the $2/3$ charged quarks. This distinguishing
feature puts a high premium on being able to detect
the charm anti-charm quark decay modes as well as the bottom
anti-bottom quark decays of Higgs particles at the LHC. It should also be
possible to calculate the contribution of the bound state to the
$\rho$ parameter, but this seems to be rather difficult in
practice.

At present the strongest evidence in favour of our scenario is
that the experimental top quark Yukawa coupling constant is,
within the crude accuracy of our calculations, in agreement with
the value at which the phase transition between the two vacua
should take place. If this agreement should persist with a more
accurate calculation of the phase transition coupling, it would
provide strong evidence in support of our scenario

There is, of course, a need for some physical mechanism underlying
our model, which could be responsible for the needed fine-tuning.
It seems likely that some kind of non-locality, through space-time
foam or otherwise, is needed \cite{glasgowbrioni}

\section*{Acknowledgements}

We should like to thank L.~Laperashvili and L.~Susskind for useful 
discussions.


\begin{thebibliography}{99}
\bibitem{glasgowbrioni}
D.L.~Bennett, C.D.~Froggatt and H.B.~Nielsen, Proceedings of the
27th International Conference on High Energy Physics, p. 557,  ed.
P. Bussey and I. Knowles (IOP Publishing Ltd, 1995); Perspectives
in Particle Physics '94, p. 255, ed. D. Klabu\u{c}ar, I. Picek and
D. Tadi\'{c} (World Scientific, 1995) [arXiv:hep-ph/9504294].
\bibitem{guendelmann}
E.I.~Guendelman, Mod. Phys. Lett. A {\bf 14}, 1043 (1999) 
[arXiv:gr-qc/9901017].
\bibitem{woodard}
N.C.~Tsamis and R.P.~Woodard, Ann. Phys. {\bf 238}, 1 (1995).
\bibitem{weinberg}
S.~Weinberg, Sources and detection of dark matter and dark energy
in the universe: Proceedings, ed. D.~Cline {Springer Verlag, 2001}
[arXiv:astro-ph/0005265].
\bibitem{itep}
C.D.~Froggatt and H.B.~Nielsen, Surv. High Energy Phys. {\bf 18}, 55 
(2003) [arXiv:hep-ph/0308144].
\bibitem{fln}
C.D.~Froggatt, L.V.~Laperashvili and H.B.~Nielsen, Hierarchy
Problem and Multiple Point Principle (in preparation).
\bibitem{colin}
C.D.~Froggatt, These Proceedings [arXiv:hep-ph/0312220].
\bibitem{fn2}
C.D.~Froggatt and H.B.~Nielsen, Phys. Lett. B {\bf 368}, 96 (1996)
[arXiv:hep-ph/9511371];\\
C.D.~Froggatt, H.B.~Nielsen and Y.~Takanishi, Phys. Rev. D {\bf
64}, 113014 (2001) [arXiv:hep-ph/0104161].
\end{thebibliography}
\end{document}